\documentclass[12pt,a4paper]{article}
\usepackage{a4wide}

\usepackage{amsmath}
\usepackage{amssymb}
\numberwithin{equation}{section}

\DeclareMathOperator{\tr}{tr}

\def\Im{\mathop{\mathrm{Im}}}
\def\rank{\mathop{\mathrm{rank}}}

\def\cN{{\mathcal N}}
\def\Mpl{M_{\text{pl}}}
\def\CP{\mathbb{CP}}
\def\WCP{\mathbb{WCP}}
\def\vev#1{\langle#1\rangle}

\usepackage{graphicx}

\begin{document}

\begin{flushright}
arXiv:0706.2114\\
UT-07-21\\
YITP-07-35
\end{flushright}
\begin{center}
{\Large\bfseries 
Rigid Limit in $\cN=2$ Supergravity\\[2mm]
and Weak-Gravity Conjecture}\\

\vskip1cm

\bigskip\bigskip\bigskip
 {\large Tohru Eguchi$^{\dagger}$ and Yuji Tachikawa$^\ddagger$}\\
\bigskip
\itshape
$^\dagger$Department of Physics, University of Tokyo, \\
Tokyo 113-0033, Japan,\\
and\\
$^\dagger$Yukawa Institute for Theoretical Physics, Kyoto University, \\
Kyoto 606-8502, Japan\\
and\\
$^\ddagger$School of Natural Sciences, Institute for Advanced Study, \\
Princeton, New Jersey 08540, USA
\end{center}

\vskip2cm

\centerline{\large\textbf{abstract}}

\vskip1cm 

We analyze the coupled $\cN=2$ supergravity and
Yang-Mills system 
using holomorphy, near the rigid
limit where the former decouples from the latter.
We find that there appears generically a new mass scale 
around $g\Mpl$ where $g$ is the gauge coupling constant
and $\Mpl$ is the Planck scale. This  is in accord with the
weak-gravity conjecture proposed recently.  We also study the scale dependence
of the gauge theory prepotential from its embedding into supergravity.

\newpage

\section{Introduction}
Quantization of general relativity has been one of the most serious challenges for  theoretical 
physics for a long time.  Its coupling constant is dimensionful,
which makes the theory apparently non-renormalizable. Thus, we need to complete the 
theory in the ultraviolet (UV) to make it into a consistent quantum theory.
The prime candidate for quantized gravity is the superstring theory, and
the progress we made during the last decade 
makes us confident that
there exist many consistent four-dimensional theories with a high degree of supersymmetry 
containing quantized graviton in their spectrum. 
These low energy field theories coupled to gravity have a consistent UV completion and are obtained via compactification of
superstring theory on suitable internal manifolds.

When we come to theories with  a  smaller number of supersymmetries the situation becomes 
somewhat delicate.  Recent developments suggest
that there exists an enormous number of $\cN=1$ supersymmetric
four-dimensional models with negative cosmological constant
(for a review, see e.g.~\cite{landscapereview}).
This {\em landscape} of superstring vacua, if taken at face value,
predicts a disturbingly huge number, $10^{200}$ or larger, of solutions
with varying gauge groups and matter contents. 
Then it is natural  to ask which theory
is realized as a low-energy effective description of
a consistent theory with quantized gravity \cite{swampland}.
Several criteria have been already proposed in
\cite{weakgravity,oogurivafa} which characterize models in the {\em swampland} 
which cannot be UV completed to a consistent theory of quantum gravity.

The criterion we will focus in this article 
is  {\em the weak-gravity conjecture} proposed in \cite{weakgravity}; 
one way to state the conjecture is that if a consistent theory coupled to 
gravity with the Planck scale $\Mpl$ contains a gauge field with the coupling
constant $g$, then there should necessarily be a new physics
around the mass scale $g\Mpl$. We refer the reader to the original article
for the arguments which led to this proposal \cite{weakgravity}.  
Our objective in this article is to show how this conjecture will  
generically hold  within the framework of 
$\cN=2$ supersymmetric Yang-Mills
coupled to $\cN=2$ supergravity. 

The system of $\cN=2$ supersymmetry is well suited to the analysis of the effects
of quantum gravity on the gauge theory.  One 
advantage is that 
the dynamics of $\cN=2$ supersymmetric Yang-Mills theories
has been studied in great detail since the pioneering work of \cite{SW}.
Another advantage is that the limit where the $\cN=2$ 
Yang-Mills theory decouples from the $\cN=2$ supergravity
is fairly well understood in the context of the string compactification on 
Calabi-Yau (CY) manifold with a fiber 
of ADE singularities.  This limit is known as
the {\em rigid limit} or {\em decoupling limit} since supersymmetry becomes rigid and 
 gravity decouples from the gauge theory in the limit. It is also called  
 the {\em geometric-engineering limit} \cite{Kachru-Vafa,selfdual,ge}, 
since non-Abelian gauge symmetry 
is generated by ADE singularities.

In this paper we consider a type II string theory on CY manifolds which possess $K3$ fibration over ${\CP}^1$ and thus has a dual heterotic string description. At the geometric engineering limit $\epsilon\rightarrow 0$ when the $K3$ surface develops ADE singularity, such a CY manifold acquires periods which behave as a power and logarithm of $\epsilon$. We  shall show that the ratio of these periods leads to the hierarchy of gauge and gravity mass scales which has exactly the form of the weak gravity conjecture. Since the geometric engineering limit is the only way to generate non-Abelian gauge symmetry in type II  theory,
the weak gravity conjecture seems to hold generically in $\cN=2$ gauge theory coupled to $\cN=2$ supergravity. Actually as is well-known, $M_{\text{het}}$=$g\Mpl$ is the mass scale of heterotic string theory and thus the weak gravity conjecture seems to fit very nicely with the type II-heterotic duality.

In our analysis the holomorphy and the special geometry of $\cN=2$ theories play the basic role.
Holomorphic functions are determined by their behavior at the singularities, in particular by the monodromy properties around the singular locus.

The organization of the paper is as follows:  In Section \ref{example}
we discuss an example of a type II string theory  compactified on 
a CY  manifold with a $K3$ fibration. We shall show how
a hierarchy of mass scales is generated in the rigid limit $\epsilon\rightarrow 0$ 
which fits exactly to the weak gravity conjecture.
We also point out that  the presence of a logarithmic period $\log \epsilon$ 
predicts a kinetic term for a field S  \begin{equation}
\frac{\partial_\mu S\partial_\mu S}{(\Im S)^2}.
\label{dilaton}\end{equation}
$S$ corresponds to the 
gauge coupling constant $S=\theta/(2\pi )+4\pi i/g^2$ and maps to the heterotic dilaton  under the 
type II/heterotic duality.  In Section \ref{generalization}
we will discuss generalization of the weak gravity hypothesis. 
We discuss in Section \ref{construction}  
the mechanism of how the logarithmic periods necessarily 
appear in a CY manifold with a $K3$ fibration.
In Section \ref{RGflow} we use the embedding of gauge theory 
into supergravity and derive the {\em renormalization group} formula for the 
dependence of the prepotential $F_{\text{gauge}}$
on the dynamical scale $\Lambda$ \cite{Matone,STY,EY}. We derive for any gauge theory of ADE group, a relation \begin{equation}
\frac{\partial F_{\text{gauge}}}{\partial \log\Lambda}={h \over \pi i} u_2.
\label{rg-eq}\end{equation} Here $u_2=\vev{\tr \phi^2}$ and 
$\phi$ denotes the adjoint scalar 
in the vector multiplet. $h$ is the Coxeter number of the group.
We conclude this note with some discussions in Section \ref{discussion}.

\section{An Example}\label{example}
\subsection{A Calabi-Yau and the rigid limit}\label{CalabiYau}
Let us start with an example from the string theory.
As is well-known, in the type IIA superstring theory, an $\cN=2$ supergravity
system in four dimensions can be  obtained by compactification
on a CY manifold $M$. It is also known that the $SU(n)$ $\cN=2$
gauge symmetry arises if $M$ has a sphere of $A_{n-1}$ type singularities. In the simplest case of 
$A_1$ singularity such a CY manifold has at least two K\"ahler parameters: 
one for the size of the sphere of the singularities,
and the other for the size of resolution of  singularities.  One explicit example
is given by a CY manifold $X_8$ which is a degree 8 hypersurface 
in the weighted projective space $\WCP_{1,1,2,2,2}^4$ with Hodge
 numbers $h_{11}=2$, $h_{21}=86$. 
 
Our analysis is facilitated by going to the mirror type IIB theory where world-sheet instanton corrections in IIA theory are summed up by mirror transformation. 
Mirror pair of $X_8$ and $X_8^*$ has been extensively studied in the literature
(e.g.~\cite{twoparameter,hosono1,explicit}). 
We first briefly review their properties. 
Defining equation of the mirror $X_8^*$ is given by 
\begin{equation}
X_8^*:\,\,W=\frac{B}{8}x_1^8+\frac{B}{8} x_2^8+\frac14 x_3^4+\frac14 x_4^4+\frac14 x_5^4 -\psi_0x_1x_2x_3x_4x_5-\frac14\psi_2(x_1x_2)^4=0 \label{CY=M}
\end{equation}
in an orbifold of  $\WCP_{1,1,2,2,2}^4$.  $[B:\psi_0:\psi_2]$
parametrizes the complex structure moduli of $X_8^*$.
We first note that  this hypersurface has a structure of a $K3$ fibration over $\CP^1$: by a change of variables $x_0=x_1x_2$, $\zeta={x_1/ x_2}$, $W$ is rewritten as
\begin{eqnarray}
&&W={B'\over 4}x_0^4+\frac14 x_3^4+\frac14x_4^4+\frac14 x_5^4-\psi_0x_0x_3x_4x_5=0,\label{K3}\\
&&B'={B\over 2}(\zeta+{1\over \zeta})-\psi_2.
\end{eqnarray}
$\zeta$ parametrizes the base of the $K3$ fibration.
$K3$ surface (\ref{K3}) (with fixed $\zeta$) has singularities at
\begin{eqnarray}
&& B'=0; \hskip7mm \mbox{large complex structure limit},\label{lcs}\\
 && B'=\psi_0^4;         \hskip5mm \mbox{conifold singularity}. \label{conifold}
\end{eqnarray}
These are located by imposing equations $W=0,\,\partial W/\partial x_i=0,\,i=0,3,4,5$ simultaneously.
If we solve (\ref{lcs}), (\ref{conifold}) for $\zeta$, we find   
\begin{eqnarray}
B'=0 &\Longrightarrow&  \zeta=e^{\pm}_0,\hskip2mm \mbox{where} \hskip2mm e^{\pm}_0={\psi_2\over B}\pm \sqrt{\left({\psi_2\over B}\right)^2-1},\label{e_0}\\
B'=\psi_0^4 &\Longrightarrow& \zeta=e^{\pm}_1,\hskip2mm \mbox{where} \hskip2mm e^{\pm}_1={(\psi_2+\psi_0^4)\over B}\pm \sqrt{\left({\psi_2+\psi_0^4\over B}\right)^2-1}.\label{e_1}
\end{eqnarray}
 
 Singularities of the total space $X_8^*$ are located by further imposing ${\partial B'/ \partial \zeta}=0$
\begin{equation}
{\partial B'\over \partial \zeta}\Longrightarrow \hskip2mm B=0 \hskip3mm \mbox{or} \hskip3mm \zeta=\pm 1.
\end{equation}
Substituting $\zeta=\pm 1$ into (\ref{lcs}),\,(\ref{conifold}) we find 
singular loci in the moduli space of $X_8^*$ 
\begin{equation}
B=\pm\psi_2,\hskip3mm B=\pm(\psi_2+\psi_0^4).
\end{equation}
These coincide with the locations where $e_0^{\pm},\,e_1^{\pm}$ become degenerate.

\begin{figure}
\centerline{\includegraphics[width=.6\textwidth]{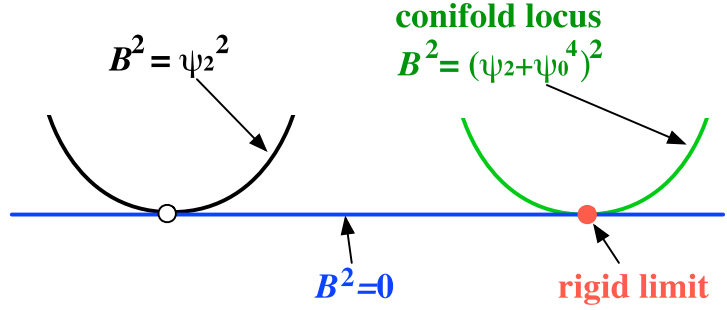}}
\caption{Discriminant loci of the moduli of the CY $X_8^*$,
before the blowup.\label{before}}
\end{figure}

Thus the discriminant of the mirror CY manifold is given by
 \begin{equation}
\Delta=B^2(B^2-\psi_2^2)(B^2-(\psi_2+\psi_0^4)^2).
\end{equation} Three components of the discriminant loci are depicted in Figure~\ref{before}.
The first and the second factor intersect tangentially at the 
large complex structure point, and the third factor is the conifold locus.
The conifold locus and the locus $B^2=0$ also meet tangentially at the rigid limit \footnote{
The parameter sets  $(B,\psi_0,\psi_2)$ and $(-B,\psi_0,\psi_2)$
describe the same complex structure, and so the natural coordinate
of the moduli is $B^2$
rather than $B$.},
so that the moduli space needs to be blown up at these points.

\begin{figure}
\centerline{\includegraphics[width=.6\textwidth]{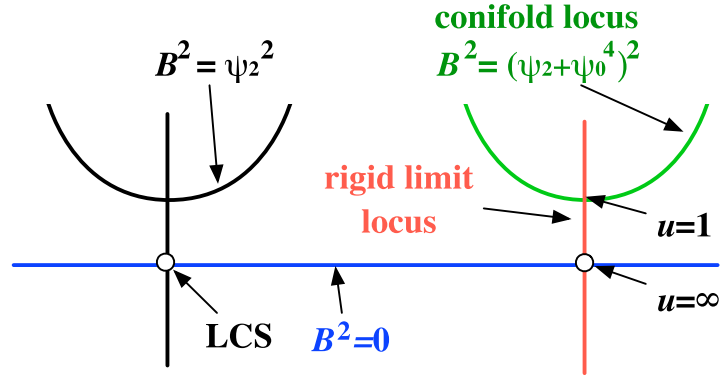}}
\caption{Discriminant loci of the moduli of the CY $X_8^*$
after the blowup. LCS stands for the Large Complex Structure point.\label{after}}
\end{figure}

We now concentrate on the region near the rigid limit.
The blowing up introduces an exceptional curve
which is a $\CP^1$ parametrized by $[\Lambda^2:u]$
via the relation
\begin{equation}
\epsilon\Lambda^2=B,\qquad \epsilon u = \psi_2+\psi_0^4. \label{blowup}
\end{equation}
The exceptional curve is at $\epsilon=0$.
The discriminant loci
after the blowup are shown in Figure~\ref{after}.

The defining polynomial $W$ 
in the limit $\epsilon\to0$ is given by
\begin{equation}
W=\frac{\epsilon}2\left[ \frac12(w+\frac{\Lambda^4}{w})
+x^2+y^2+z^2-u\right]+O(\epsilon^{2}).
\label{SWgeometry}
\end{equation} after a suitable redefinition of the coordinates.
 This is a fibration of $A_1$ singularity over
$\CP^1$ parametrized by $w$. It is in fact the
Seiberg-Witten geometry of the $\cN=2$ supersymmetric
pure $SU(2)$ Yang-Mills theory with the modulus
$u=\vev{\tr\phi^2}$ and the dynamical mass scale $\Lambda$.
Thus, the exceptional curve we have introduced is identified
as the $u$-plane of $SU(2)$ gauge theory:
the $u$-plane is naturally compactified at $u=\infty$ into a sphere. 
We call this sphere
the rigid limit locus.   

Note that  before taking the rigid limit $\epsilon\rightarrow 0$,
 the theory contains $h_{11}+1=3$ 
gauge fields: they are the graviphoton, the gauge partner of 
the scalar field $S$ and the $U(1)$ (Cartan-subalgebra) part of $SU(2)$ gauge field. 
Here $S$ denotes the scalar field which corresponds to the gauge coupling constant 
in field theory,
\begin{equation}
S={\theta\over 2\pi}+{4\pi i\over g^2}.
\end{equation} 
We recall that when CY manifold $M$ possesses a $K3$ fibration on $\CP^1$, there exist a duality between 
type IIA on $M$ and heterotic theory on $K3\times T^2$ \cite{AL}.  
The field $S$ corresponds to the size of the base 
$\CP^1$ of $K3$ fibration in type IIA theory and becomes the heterotic dilaton under this duality.
 In the decoupling limit $\epsilon \rightarrow 0$,
  two of the gauge fields, the graviphoton 
and the partner of $S$, 
disappear and we are only left with the (Cartan part of) $SU(2)$ gauge field.

\subsection{Behavior of the K\"ahler potential}
Let us next quickly recall the structure of vector multiplet
scalars in the $\cN=2$ theories.  First, in the case of field theories 
of rigid $\cN=2$ supersymmetry with the gauge group $U(1)^n$,
there exist $n$ complex scalar fields $\phi^i$, $(i=1,\ldots,n)$. 
Their K\"ahler potential is given by \begin{equation}
K=\Im \sum_i(a^D_i)^* a^i
\label{kahler-gauge}\end{equation} where
$a^i$ and $a^D_i$ are holomorphic functions of the VEV's 
of $\phi^i$.
 $a^i$ and $a^D_i$ are  called the special coordinates
or the periods of the theory. Dual periods are 
related to each other as \begin{equation}
a^D_i=\frac{\partial F_{\text{gauge}}}{\partial a^i}, \hskip3mm i=1,\cdots,n
\end{equation}
where $F_{\text{gauge}}$ denotes the prepotential.

 Secondly, in the case of  $\cN=2$
supergravity with $N$ vector multiplets, 
there exist $2(N+1)$ periods $X^a,F_a$, $a=1,\cdots,N+1$. 
The K\"ahler potential is given by \begin{equation}
e^{-K}=\Im\sum_a F_a^* X^a.
\label{sugra-K}\end{equation}  
The periods $X^a$, $F_a$ are holomorphic functions of scalars $\Phi^i$, $(i=1,\ldots,N)$.
Under the K\"ahler transformation $K\to K-f-f^*$ 
periods are transformed as $X^a\to e^f X^a$, $F_a\to e^f F_a$.
The mass squared of a BPS-saturated soliton with charges $(q_a,m^a)$
is then given by \begin{equation}
m^2=e^{K} |\sum_a(q_a X^a + m^a F_a)|^2,\label{sugraBPS}
\end{equation}which is invariant under the K\"ahler transformation.
An important property of the supergravity periods 
is the transversality condition: \begin{equation}
\sum_a X^a \frac{\partial F_a}{\partial\Phi^i}-\sum_a \frac{\partial X^a}{\partial\Phi^i} F_a=0,
\label{transversality}
\end{equation} which guarantees the existence of the prepotential.
Prepotential of $\cN=2$ supergravity is a homogeneous function of degree 2 in $X_a$. 

In the case of CY compactification of type IIB string theory,
the periods are given by \begin{equation}
X^a=\int_{A^a}\Omega,\qquad F_a=\int_{B_a}\Omega
\end{equation}where $\Omega$ is the $(3,0)$-form of the CY
and $A^a$, $B_a$ are the canonical basis of $H_3(M^*,\mathbb{Z})$ of CY manifold. 
In this case 
the condition \eqref{transversality}
comes from the Griffiths transversality
 $\int \Omega\wedge \partial_{\Phi^i}\Omega=0$.

Now let us go back to the example of the previous section, type IIB string theory compactified on $X_8$.
In the field theory limit we have only one gauge field ($n=1$) and two periods 
$a$ and $a^D$ of $SU(2)$ Seiberg-Witten theory.
At the level of supergravity there exist three gauge fields (two vector multiplets, $N=2$) and 
six periods $X^a,F_a,\,a=1,2,3$.
Behavior of these periods near the decoupling limit and in particular their monodromy properties 
around rigid limit locus have been discussed in great detail in \cite{explicit}.

It turns out that two of the periods, say $X^1$ and $F_1$, are converted to the gauge theory periods in 
the rigid limit. They behave as
 \begin{equation}
X^1=\epsilon^{1/2} a + O(\epsilon),\qquad
F_1=\epsilon^{1/2} a^D + O(\epsilon).
\end{equation}
Remaining four periods behave as
\begin{equation}
X^2,\,X^3=1+O(\epsilon^{1/2}),\qquad
F_2,\,F_3=\frac{1}{2\pi i}\log\epsilon + O(1).
\end{equation}
The origin of logarithmic behaviors in $F_{2},F_3$ will be discussed 
in Section \ref{construction}:
they come from the geometry of $K3$ fibration of the CY manifold.
 
 Then using (\ref{sugra-K}) we find that $e^K$ behaves as $\log 1/|\epsilon|$.
Therefore the supergravity K\"ahler potential is expanded as
\begin{equation}
K=\log (\log 1/|\epsilon|) + \frac{|\epsilon|}{\log1/|\epsilon|}
\Im (a^D)^* a + \cdots.\label{Kaehler}
\end{equation} as $\epsilon\to 0$.  
Note that $\Im\left(a^D\right)^*a$ 
is the K\"ahler potential of the field theory (\ref{kahler-gauge}).
Thus we can clearly see
that $SU(2)$ super Yang-Mills theory decouples from
gravity.

The factor $|\epsilon|$ in front of the K\"ahler potential of the field theory
determines the hierarchy between the Planck scale and the scale of the gauge theory:
it is basically in accord with the expectation \cite{ge} with 
$|\epsilon|^{1/2}$ being identified with the dynamical mass scale 
$\Lambda_{\text{gauge}}$ of the gauge theory.
The existence of an extra factor of $\log1/|\epsilon|$
in the denominator 
was first recognized by the authors of \cite{explicit}.
We will see in the following that this factor implies 
the weak-gravity conjecture in the present context. 

Let us now consider the weak coupling region of gauge theory for the sake of simplicity.
There the periods $a$ and $a^D$  behave as
\begin{equation}
a\approx \sqrt{2u},\qquad a^D\approx {i\over \pi}\sqrt{2u}\log u.
\end{equation}
Using the relation of periods to the low-energy gauge coupling 
constant $\tau$ : \begin{equation}
\tau=\frac{\theta}{2\pi} + \frac{4\pi i}{g^2(m_W)} =
\frac{\partial a^D}{\partial a},
\end{equation}
we find \begin{equation}
e^{-2\pi^2/g^2(m_W)} =u^{-1/2}.\label{A}
\end{equation} The coupling constant $g$ in the above equation is to be evaluated
at the scale of the mass  $m_W$ of the massive gauge boson
where the coupling stops running.
$m_W$ is, in turn, given by the formula
\eqref{sugraBPS}  \begin{equation}
m_W^2= e^{K}|X^1|^2 =
\frac{|\epsilon|}{\log 1/|\epsilon|}u.\label{B}
\end{equation} From \eqref{A} and \eqref{B}, we find the  dynamical scale of the
gauge theory  \begin{equation}
\Lambda_{\text{gauge}}=m_W e^{-2\pi^2/g^2(m_W)}=
\frac{|\epsilon|^{1/2}}{(\log1/|\epsilon|)^{1/2}} \Mpl \label{xxx}
\end{equation} where we reinstated the Planck scale to recover the correct mass dimension.

Let us next introduce a chiral superfield $S=\theta/2\pi+4\pi i/g^2$
via the relation \begin{equation}
S=\frac{1}{\pi i }\log \epsilon.
\label{S-approx}\end{equation} 
Then, the monodromy around $\epsilon=0$ is generated by
the shift $S\to S+2$. 
$\Im S$, which is  the partner of the dynamical
theta angle, is the natural bare gauge coupling constant in the supergravity.
Furthermore, $S$ coincides with the heterotic dilaton which we have discussed at the end of Section \ref{CalabiYau}. 
There will be subleading corrections to 
(\ref{S-approx}) if one goes outside the region of weak coupling 
or small $\epsilon$.
Another notable fact is that, because of the K\"ahler potential \eqref{Kaehler}, 
the field $S$ in  fact has the standard kinetic term for the dilaton, \begin{equation}
g_{SS^*}\partial_\mu S\partial_\mu S^*=
\frac{\partial_\mu S\partial_\mu S^*}{(\Im S)^2}.
\end{equation}  

\begin{figure}
\centerline{\includegraphics[width=.5\textwidth]{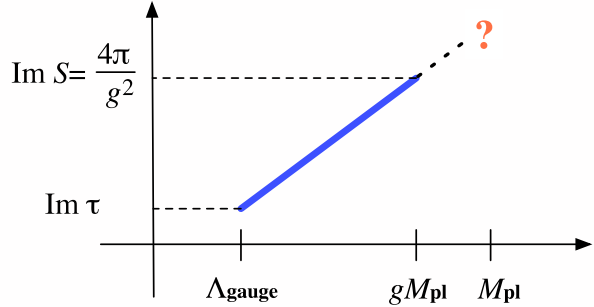}}
\caption{Running of the  coupling in the gauge theory coupled to supergravity. \label{running}}
\end{figure}

Using the field $S=\theta/2\pi+4\pi i/g^2$,
the relation \eqref{xxx} now becomes \begin{equation}
\Lambda_{\text{gauge}}=e^{-2\pi^2/g^2}\,\cdot \, g \Mpl.
\end{equation} There exists an extra factor of $g$ in front of $\Mpl$ in the above equation, which means that the ultraviolet gauge coupling $g$ is defined 
not at the Planck scale $\Mpl$ but at a lower energy scale $g\Mpl$.
The running of the gauge coupling from the value at low energy $\Im\tau$
to the one at high energy $\Im S$ 
is schematically depicted in Figure~\ref{running}.
The existence of the new scale $g \Mpl$ is what the weak gravity conjecture
has predicted. Thus  the analysis of the $\cN=2$
$SU(2)$ gauge theory coupled to supergravity supports the weak gravity conjecture.

\section{Generalization}\label{generalization}
Let us consider what happens in the generic 
$\cN=2$ gauge theory coupled to $\cN=2$ supergravity.  
Suppose the gauge theory has $n$ vector multiplets.
In the coupled gauge-gravity system the gauge coupling constant is promoted to 
a scalar field $S$. Thus there is at least one extra
vector multiplet in the locally supersymmetric theory. 
Let us consider the minimal situation; i.e. the total number
of the $U(1)$ vector multiplets being equal to $n+1$. 
Then, altogether there are $n+2$ gauge fields including the graviphoton 
and  there will be
$2n+4$ supergravity periods. Therefore, by coupling the gauge theory
to the $\cN=2$ gravity  we should have at least four extra periods.

We assume that there is a locus $E$
in the suitably blown-up moduli space given by the local parameter
$\epsilon=0$ around which  some $\Omega_i$, $(i=1,\ldots,2n)$
of the periods $\Omega_I$, $(I=1,\ldots,2(n+2))$ become
parametrically small, $\Omega_i \propto O(\epsilon^{1/h})$
for some power $h$. 
This statement itself is not invariant under K\"ahler transformation,
so we also demand that there will be periods which stay constant near $E$. 

The monodromy around $E$ may also be logarithmic: thus there might be
periods behaving as $\propto (\log \epsilon)^k$. Let $p$ be the largest 
power $k$ of such periods. 
There is a mathematical theorem\footnote{see e.g.~Appendix A of \cite{ET}
and references therein.}
which then states the K\"ahler potential behaves as \begin{equation}
e^{-K}= \Im \sum_a F^*_a X^a \propto (\log |\epsilon| )^p.
\end{equation}

Let us define the chiral field $S$ by \begin{equation}
S=\frac1{\pi i}\log\epsilon=\frac{\theta}{2\pi}+\frac{4\pi i}{g^2}
\end{equation}  as before.
Repeating the argument presented in the last section, we readily obtain
a relation \begin{equation}
\Lambda_{\text{gauge}} \sim e^{-4\pi^2/(h g^2)}\,\cdot\, g^p\Mpl.
\end{equation} 
We will see in the next section that
$h$ equals  the quadratic Casimir of the gauge group
in the case of pure $\cN=2$ gauge theory.

Furthermore, the kinetic term of $S$ 
is given by  \begin{equation}
\frac{\partial_\mu S \partial_\mu S^*}{(\Im S)^2}\qquad
\text{or}\qquad
\partial_\mu S \partial_\mu S^*
\end{equation} depending on  $p\ne 0$ or $p=0$, respectively.
Thus, the weak gravity conjecture in $\cN=2$
supergravity coupled to super Yang-Mills follows from the existence
of a logarithmic period $\sim (\log \epsilon)^p$, $p\ge 1$. Furthermore,
the appearance of such logarithmic periods is related to the
field $S=\log \epsilon$ corresponding to the dilaton in heterotic theory.

\section{Explicit description of logarithmic periods}\label{construction}

For a CY which is a $K3$ fibration over $\CP^1$, 3-cycles can be constructed
explicitly. We follow the approach of \cite{explicit} and  Appendix in \cite{DenefWeak} .
Consider a CY with a defining equation \begin{equation}
w+\frac{\mu^2}w + W_{K3}(x,y,z;t_\ell)=0\label{CYeq}
\end{equation} where $t_\ell$ denote the moduli of the $K3$. 
The holomorphic 3-form is given by \begin{equation}
\Omega=\frac {dw}{w}\wedge \Omega_{K3},\qquad
\Omega_{K3}=\frac{dx\wedge dy}{\partial_z W_{K3}}.
\end{equation}   3-cycles of CY are made of the product
of a 1-cycle of the $\CP^1$ base and  a 2-cycle of $K3$. 

2-cycles of $K3$ to be used here are those which are not holomorphically embedded into $K3$, since 
holomorphic cycles have the representative which are of the $(1,1)$-form so that their integrals with the  
(2,0)-form $\Omega_{K3}$ must vanish. 
Holomorphic cycles of $K3$ form the Picard lattice $\Lambda^{\text{Pic}}$ 
\begin{equation}
\Lambda^{\text{Pic}}=H^{1,1}(K3)\cap H^2(K3,\mathbb{Z})
\end{equation}
and its dimension is called 
the Picard number $\rho(K3)$. Cycles which are not holomorphically embedded are called transcendental and the lattice $\Lambda$ of the 2nd homology of $K3$ has an orthogonal decomposition into Picard and transcendental lattices 
\begin{equation}
\Lambda=\Lambda^{\text{Pic}}\oplus\Lambda^{\text{tr}}.
\end{equation}
It is well-known that the lattice $\Lambda$ has a signature of $(3,19)$. 
In the case of projective $K3$, the K\"ahler form becomes algebraic and the Picard lattice has a signature $(1,\rho(K3)-1)$.  
Then the signature of $\Lambda^{\text{tr}}$ becomes $(2,20-\rho(K3))$. 

In the case of the quartic $K3$ surfaces which featured in our example $X_8$, the Picard number is $\rho(K3)=19$ and thus there are three transcendental cycles with signature $(2,1)$.  The 2-cycle with a negative signature, i.e. a negative self-intersection number 
is the vanishing cycle of $A_1$ singularity. 
Two 2-cycles of the positive signature 
generate periods which have logarithmic behavior in $\epsilon$ 
as we see below.  Ref.~\cite{explicit} discusses another example of CY manifold $X_{24}$ which also possesses a $K3$ fibration and produces 
the $SU(3)$ gauge theory in the decoupling limit. In this case there exist four transcendental cycles with a signature $(2,2)$. Two 2-cycles with the negative signature describe the vanishing cycles of $A_2$ singularity. In the case of general $A_r$ singularity there will be $2+r$ transcendental cycles with the signature $(2,r)$. As we shall see below, 
two transcendental cycles of $K3$ with the positive 
signature will generate logarithmic cycles of CY manifold. 
\begin{figure}
\centerline{\includegraphics[height=.3\textwidth]{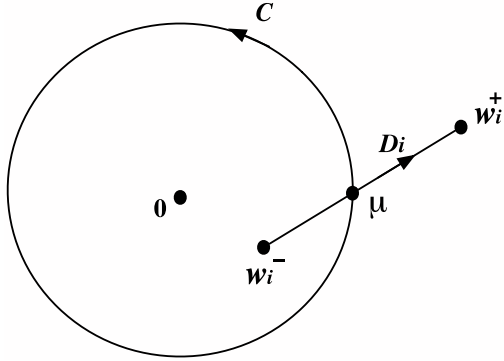}}
\caption{Cuts in the base $\CP^1$.\label{paths}}
\end{figure}

\begin{table}\[
\begin{array}{l|l|c|ccc|l}
&\text{defining eq.}&h&d_x&d_y&d_z&\text{deg.~of Casimirs}\\
\hline
A_{n-1}&0=x^2+y^2+z^n&n&n/2&n/2&1 &2,3,\ldots,n\\
D_{n+1}&0=x^2+y^2z+z^n&2n&n&n-1&2&2,4,\ldots,2n\\
E_6&0=x^2+y^3+z^4&12&6&4&3&2,5,6,8,9,12\\
E_7&0=x^2+y^3+yz^3&18&9&6&4&2,6,8,10,12,14,18\\
E_8&0=x^2+y^3+z^5&30&15&10&6&2,8,12,14,18,20,24,30
\end{array}
\]
\caption{Data of ADE singularities. \label{ADE}}
\end{table}

The CY \eqref{CYeq}
 can be thought of as a one-parameter family of $K3$, whose moduli depend
on  $w$.  Suppose a transcendental two-cycle $S_i$ degenerates 
at $w+\mu^2/w=k_i$.  For a small $\mu$,  this happens
at $w_i^+\sim k_i$ and $w_i^-\sim \mu^2/k_i$, see Figure~\ref{paths}.
Let $C$ be the circle around the origin $|w|=|\mu|$, and $D_i$ denote the path
connecting $w_i^{\pm}$.  Then $C\times S_i$ and $D_i\times S_i$ are
closed 3-cycles of CY manifold.

In general, Yang-Mills gauge theories are geometrically engineered 
by fine-tuning  the parameters $\{t_{\ell}\}$ of $K3$ 
so that the $K3$ develops ADE singularities: see \cite{selfdual} for $SU(n)$,
\cite{AG} for $SO(n)$ and \cite{Brodie,HT} for $E_n$ groups.
Suppose we have a singularity of type $G$ with $\rank G=r$ around $x=y=z=0$.  
The moduli $\{t_{\ell}\}$ of $K3$ are decomposed into two sets of parameters 
 \begin{eqnarray}
\{u_2,\cdots,u_h\} ,\hskip3mm 
\{v_1,v_2,\cdots\},
\end{eqnarray} where $u_i$ corresponds to the degree $i$  Casimir invariant
 of the group $G$. $u_i$ are tuned to vanish as $\epsilon^{i/h}$ in the geometric 
 engineering limit and we rescale 
 them as $\epsilon^{i/h}\cdot u_i$. Here $h$ is the dual Coxeter number of $G$. $v_j$ are the moduli which remain finite 
in the engineering limit.
 We also introduce the rescaled coordinates  as
 \begin{equation}
w=\epsilon \tilde w,\quad x=\epsilon^{d_x/h} \tilde x,\quad
y=\epsilon^{d_y/h} \tilde y, \quad
z=\epsilon^{d_z/h} \tilde z.
\end{equation} $d_{x,y,z}$ are the degrees of $x,y,z$ (see Table~\ref{ADE}). 
We also set $\mu=\epsilon \Lambda^h$.
Then the defining equation \eqref{CYeq} of the CY
becomes \begin{equation}
\epsilon\left(\tilde w+\frac{\Lambda^{2h}}{\tilde w}+
W_{ADE}(\tilde x,\tilde y,\tilde z;u_i)+O(\epsilon^{1/h})\right)=0.
\end{equation} The holomorphic 3-form is given by \begin{equation}
\Omega = \frac{dw}{w}\wedge \frac{dx\wedge dy}{\partial_z W_{K3}}
= \epsilon^{(d_x+d_y+d_z)/h-1}\frac{d\tilde w}{\tilde w}\wedge \frac{d\tilde x\wedge d\tilde y}
{\partial_{\tilde z} W_{ADE}}
=\epsilon^{1/h}\frac{d\tilde w}{\tilde w}\wedge\Omega_{ADE}
\end{equation} where we used the fact $d_x+d_y+d_z=h+1$.

There are $r$ independent two-cycles $S_i$ of $K3$ which vanish simultaneously 
in the engineering limit.
These give rise to $2r$ 3-cycles $\bar A_i=C\times S_i $ and $\bar B_i=D_i\times S_i,\,i=1,\cdots,r$ 
of the CY as explained above. 
We can take their linear combinations, $A^i$ and $B_i$,
so that they have the canonical  intersection form,
$(A^i,A^j)=(B_i,B_j)=0$, $(A^i,B_j)=\delta^i_j$.
Then \begin{equation}
a^i =\int_{A^i} \frac{d\tilde w}{\tilde w} \wedge \Omega_{ADE},\qquad
a^D_i =\int_{B_i} \frac{d\tilde w}{\tilde w} \wedge \Omega_{ADE},\qquad
\end{equation} are identified with the special coordinates
of Seiberg-Witten theory.
Corresponding supergravity periods behave as \begin{equation}
X^i=\int_{A^i}\Omega = \epsilon^{1/h} a^i + O(\epsilon^{2/h}),\qquad
F_i=\int_{B_i}\Omega = \epsilon^{1/h} a^D_i + O(\epsilon^{2/h}).\label{rigidperiods}
\end{equation}

$K3$ surface has two extra 2-cycles which have a positive signature,
 as we have noted above.
We call them $T_a,\,a=1,2$ and arrange them so that they do not intersect $S_i$
and stay at finite values of $x$, $y$ and $z$.  
Now   the defining equation of CY  
near the 3-cycles $T_a$ is given by 
  \begin{equation}
w+\frac{\epsilon^{2} \Lambda^{2h}}w + W_{K3}(x,y,z;0, v_j)=0
\label{def-CY}\end{equation} 
Thus from the cycle  $U_a=C\times T_a$ we obtain the period \begin{equation}
\Omega_{U_a}=\int_{U_a} \Omega = \oint _C{dw \over w} 
\int_{T_a}\Omega_{K3} = 2\pi i \,  c_a\approx O(1) \quad \text{where}
\quad c_a= \int_{T_a} \Omega_{K3}(u_i=0;v_j).
\end{equation} 
In the case of the cycles $V_a=D_a\times T_a$, 
the end points of the $w$ integration become
\begin{equation}
 w_a^-\sim { \epsilon^2\Lambda^{2h}\over k_a},\hskip3mm w_a^+\sim k_a
\end{equation} where $k_a=w+\epsilon^2\Lambda^{2h}/w$ is the value
at which the 2-cycle $T_a$ degenerates.
Then we find the logarithmic behavior \begin{equation}
\Omega_{V_a}=\int_{V_a}\Omega=
\int_{ w_a^-}^{ w_a^+} {d w \over  w} 
\int_{T_a}\Omega_{K3} \approx -2 c_a  \log{\epsilon}.
\end{equation} The analysis of the monodromy under the phase rotation of 
$\epsilon$ suggests 
\begin{equation}
\Omega_{V_a}\approx -\frac{1}{\pi i}\log \epsilon \cdot \Omega_{U_a} + O(\epsilon^{1/h}),
\end{equation} although
the precise form of this expression will depend on the intersection form of $T_a$.
Thus we have established the existence of periods behaving logarithmically 
near the engineering limit.

\section{Renormalization Group Equation}\label{RGflow}

As an application of the above analysis, we shall derive the relation \begin{equation}
\frac{\partial F_{\text{gauge}}}{\partial \log \Lambda}={h\over \pi i}u_2 \label{RG}
\end{equation} for pure $\cN=2$ Yang-Mills theory with gauge groups $G=A,D,E$
from its embedding into supergravity. 
Here $u_2=\vev{\tr\phi^2}$ is the second order Casimir
and is a monodromy-invariant
coordinate of the moduli space. 
The relation describes 
the scaling violation of the prepotential of gauge theory and is called the renormalization
group equation.

Before we start describing our derivation, let us recall how the equation \eqref{RG}
was obtained from the point of view of the gauge theory.
Originally it was derived for $SU(2)$ in \cite{Matone}
using the Picard-Fuchs equation for Seiberg-Witten curve,
and later it was generalized to the classical gauge groups in \cite{STY,EY}
using the property of the hyperelliptic curve describing the dynamics of the theory.
For the $E$-type gauge groups the relation has not been given from
the SW curve because of its complexity; 
thus our method gives the first verification of the relation for the $E$-type gauge groups.

The relation has been used in the analysis of the geometrical engineering limit
in  one of the earliest papers on the subject  \cite{Kachru-Vafa};
here instead, we derive it from the study of the periods near the engineering limit.
It was conjectured already in \cite{EY} that the relation should have
a natural interpretation in supergravity since $\log\Lambda$ 
is no longer an external parameter but becomes a VEV of a field in supergravity.

The relation  should also follow from the microscopic calculations: 
 Recall the fundamental relation in the path integral which states that \begin{equation}
\vev{\partial_\lambda L_0}=\partial_{\lambda}L_{\text{eff}} 
\end{equation}where $L_0$ is the bare Lagrangian and $L_{\text{eff}}$
is the low-energy effective Lagrangian including the quantum correction.
$\lambda$ denotes some coupling constant of the theory.
In the case of  supersymmetric theories one can likewise show \begin{equation}
\vev{\partial_\lambda W_0}=\partial_{\lambda}W_{\text{eff}},\qquad
\vev{\partial_\lambda F_0}=\partial_{\lambda}F_{\text{eff}} 
\end{equation} where $W$ and $F$ are the super and  prepotential, respectively.
Now in the $\cN=2$ theory $F_0=\tau_0 \tr \phi^2$.
Then the relation \eqref{RG} follows because 
 $\log\Lambda\propto \tau_0$  is the bare coupling constant
and $u_2=\vev{\tr{\phi^2}}$.
 It can be seen more explicitly
in the framework of multi-instanton calculation \cite{DKM}.
So the prepotential constructed from the SW curve 
 should satisfy the relation. 

Let us now turn to our derivation.
Instead of the relation \eqref{RG} itself, we shall show that its derivative with respect
to the moduli satisfies \begin{equation}
\frac{\partial^2 F_{\text{gauge}}}{\partial u_j \partial\log\Lambda} = {h \over \pi i}\,\delta^j_{2}.
\label{der}
\end{equation} Then \eqref{RG} follows by integration. The integration constant
is zero by virtue of the homogeneity of $F_{\text{gauge}}$.
The homogeneity can be used to rewrite LHS   
of \eqref{der} and rewrite it as follows:\begin{align}
\frac{\partial^2 F_{\text{gauge}}}{\partial u_j \partial\log\Lambda}
&=\frac{\partial }{\partial u_j} 
\left(2F_{\text{gauge}}- \sum_i a^i \frac{\partial F_{\text{gauge}}}{\partial a^i}\right)
=\sum_i\frac{\partial a^i}{\partial u_j}a^D_i -
\sum_i a^i\frac{\partial a^D_i}{\partial u_j}.\label{trick}
\end{align}
Now we use the relation \eqref{rigidperiods} between the periods of rigid and local theory
and  obtain \begin{equation}
\sum_i F_i \frac{\partial X^i}{\partial {u_j}}  -\sum_i X^i \frac{\partial F_i}{\partial{u_j}}
=\epsilon^{2/h}\frac{\partial^2 F_{\text{gauge}}}
{\partial u_j \partial\log\Lambda}+O(\epsilon^{3/h})
\label{expansion}
\end{equation}
One of the fundamental properties of the special geometry in supergravity
is the transversality condition \eqref{transversality}.
We decompose the periods into two sets as $(X^i,F_i; X^a,F_a)$
where $X^i,F_i$ are the periods which become those of the gauge theory 
of the rigid limit, and $X^a,F_a$ are the extra periods in supergravity.
We have\begin{equation}
\sum_i F_i \frac{\partial X^i}{\partial {u_j}}  -\sum_i X^i \frac{\partial F_i}{\partial{u_j}}
=
-\sum_a F_a \frac{\partial X^a}{\partial {u_j}} +\sum_a X^a \frac{\partial F_a}{\partial{u_j}}.
\label{transv}
\end{equation} 

As shown in the previous section, $X^a$ and $F_a$ have at most $\log \epsilon$
singularity and the rest are analytic in $\epsilon^{i/h} u_i$.
Furthermore, the logarithmic terms cancel in the RHS of \eqref{transv} 
because the LHS is analytic in $\epsilon^{1/h}$.
Therefore we have \begin{equation}
\sum_i F_i \frac{\partial X^i}{\partial {u_j}}  -\sum_i X^i \frac{\partial F_i}{\partial{u_j}}
=
-\sum_a F_a \frac{\partial X^a}{\partial {u_j}} +\sum_a X^a \frac{\partial F_a}{\partial{u_j}}
= \text{const}\cdot\epsilon^{2/h} \,\delta^j_2 +O(\epsilon^{3/h}).
\end{equation} Comparing with \eqref{expansion},
we obtain \eqref{der} up to a constant factor.

Two comments are in order: first, the constant factor is non-trivial to determine
in general but should be straightforward to fix in specific cases. It then fixes
the proportionality factor between $u_2$ entering in the geometry and $\vev{\tr\phi^2}$.
Second, the derivation above was so simple that it makes us suspicious 
why a similar analysis cannot be done in the field theory limit.
Indeed, RHS of \eqref{trick} is a monodromy-invariant quantity
of mass dimension $2-j$. Thus, it is a rational function of $u_j$'s of dimension
$2-j$ and it is forced to be $\delta^2_j$ once one can argue it does not have poles.
This is precisely the hard part because 
the special coordinates $a^i$ and $a^D_i$ are complicated functions of $u_j$'s
with a lot of cuts.
In our derivation, we utilize the fact that the extra supergravity periods
are analytic in $\epsilon^{j/h} u_j$, which does the job.

\section{Discussion}\label{discussion}
In this article, we have seen how the holomorphy inherent in
$\cN=2$ supersymmetry can be effectively 
used to study the effect of gravity upon the running of gauge theory.
More specifically, we showed how the monodromy of the periods
around the locus of the rigid limit translates to the hierarchical 
separation of the dynamical scale of gauge theory and the Planck scale. 
We have argued that, as compared to the naive relation \begin{equation}
\Lambda_{\text{gauge}} \approx e^{-4\pi^2/hg^2}\,\, \Mpl
\end{equation} there is generically an extra factor of the gauge coupling 
constant $g$ in the right hand side, \begin{equation}
\Lambda_{\text{gauge}} \approx e^{-4\pi^2/hg^2}\,\cdot\, g\Mpl
\end{equation} supporting the weak gravity conjecture.
We have also seen how the scaling violation of the prepotential of the gauge theory,
\eqref{RG}, can be naturally understood from the embedding into supergravity.

The result presented here is only a small step in utilizing
the holomorphy to understand the dynamics of the coupled $\cN=2$ 
supergravity-gauge systems.
We believe many more properties can be learned in a similar manner. 
It would also be interesting to make a comparison
with  the result in \cite{rw} where the authors calculated the one-loop effect
of gravity to the beta function of the gauge theory.  
It was argued in \cite{recent} that the beta function in \cite{rw} alone
leads to the weak gravity conjecture.  We will have to supersymmetrize
the result of \cite{rw} to carry out the comparison to our case.

It will be very important to see if it is possible to extend our results to the realm of  $\cN=1$ supersymmetric theories.
In the case when $\cN=1$ theories are obtained from those of $\cN=2$ by introducing fluxes, branes etc. 
many of the structures of the latter survive. Hopefully we will have enough control over mass scales of these theories to derive the characterization of consistent $\cN=1$ field theories coupled to gravity.

\bigskip
\section*{Acknowledgements}
The authors would like to thank Nima Arkani-Hamed and Seiji Terashima for discussions.
TE would like to thank the Institute for Advanced Study where a part of the work was done.
Research of TE is supported in part by a Grant-in-Aid from the Japan Ministry of Education and Science.  
Research of YT is supported by DOE grant DE-FG02-90ER40542.

\end{document}